\documentclass[prx,twocolumn,super,showpacs,groupedaddress,superscriptaddress]{revtex4}
\usepackage{graphicx}
\usepackage{color}
\usepackage{amsmath}
\usepackage{latexsym}
\usepackage{amssymb}

\begin{document}
\title{Stochastic sensing of polynucleotides using patterned nanopores}

\author{Jack A. Cohen}
\email[]{j.cohen@physics.ox.ac.uk}
\affiliation{Rudolf Peierls Centre for Theoretical Physics, University of Oxford, Oxford OX1 3NP, UK}

\author{Abhishek Chaudhuri}
\email[]{a.chaudhuri1@physics.ox.ac.uk}
\affiliation{Rudolf Peierls Centre for Theoretical Physics, University of Oxford, Oxford OX1 3NP, UK}
\affiliation{Department of Biomedical Science, University of Sheffield, Sheffield S10 2TN, UK}

\author{Ramin Golestanian}
\email[]{ramin.golestanian@physics.ox.ac.uk}
\affiliation{Rudolf Peierls Centre for Theoretical Physics, University of Oxford, Oxford OX1 3NP, UK}

\begin{abstract}
The effect of the microscopic structure of a pore on polymer translocation is studied using
Langevin dynamics simulation, and the consequence of introducing patterned stickiness inside
the pore is investigated. It is found that the translocation process is extremely sensitive
to the detailed structure of such patterns with faster than exponential dependence of translocation
times on the stickiness of the pore. The stochastic nature of the translocation process
leads to discernable differences between how polymers with different sequences go through
specifically patterned pores. This notion is utilized to propose a stochastic sensing protocol
for polynucleotides, and it is demonstrated that the method, which would be significantly faster
than the existing methods, could be made arbitrarily robust.
\end{abstract}

\date{\today}

\maketitle

\section{Introduction}

The quest for efficient sequencing of single stranded DNA
using synthetic nanopores has recently led to the development
of a plethora of novel theoretical and experimental design ideas that
use a variety of different approaches
\cite{Service2006,Lagerqvist2006,Branton2008,Shendure2008,Schloss2008,Zwolak2008,Min2011}.
Experiments have demonstrated that the current blockade readout from
single stranded DNA and RNA molecules that are electrophoretically
driven through biological and synthetic nanopores could in principle
reflect a signature of the underlying sequence
\cite{Kasianowicz1996, Braha1997, Akeson1999, Meller2000, Meller2001,Deamer2002, Meller2003}.
It is now possible to design solid-state nanopores \cite{Storm2003,Kim2006} with tailored
surface properties that could regulate DNA-pore surface interaction
\cite{Ohshiro2006,Iqbal2007,Wanunu2007} and also reduce noise \cite{Chen2004,Tabard2007}.
A number of recent experiments have been successful in discriminating between polynucleotides
\cite{Derrington2010} and identifying single nucleotides \cite{Clarke2009,Stoddart2009,polonsky2007}.
However, more remains to be done to resolve issues involving stability, sensitivity, and
resolution, before they can be integrated into fast and efficient devices for sequencing purposes
\cite{Hall2010,Schadt2010}.

Theoretical studies of polymer translocation through nanopores \cite{lubensky,luo2006jcp,muthukumar,matysiak,luo2007prl,luo2008pre,luo2008prl,gauthier,luan,nikoubashman,sung,muthukumar1999,muthukumar2001,kantor2001,muthukumar2003,metzler,slonkina,kantor2004,milchev2004,gerland,gopinathan,wong,milchev2011,abdolvahab2011}
have revealed that the process is intrinsically stochastic and features a rather
wide distribution for the translocation time. The inherent noise acts as an overwhelming
source of error for the sequence detection strategies that are based on deterministic
patterns in the translocation time readout, unless the process is sufficiently slowed
down such that time-averaging eliminates the noise \cite{Deamer2002,Derrington2010}. In other words,
achieving robustness in sequencing using deterministic strategies has intrinsic limitations,
and might require significant compromise in translocation speed \cite{Branton2008}.

Here, we propose a strategy to control the translocation time and its statistics
by using pores that have patterned surface energetics. We then address the question of whether
it is possible to engineer distinct stochastic features for the translocation of heteropolymers
with any given sequence through different pores, such that the statistical readout from combined
translocation events of a particular sample through a collection of different pores could
quickly and accurately reveal its sequence by synergistic exclusion of unlikely sequences.
We start by studying the translocation of a  homopolymer that is driven from the cis (entrance)
to the trans (exit) side of a narrow pore by a uniform external field, $F$ (Fig. \ref{f:sys}; see Appendix A).
We vary the stickiness of the pore (characterized by the attractive strength, $\epsilon_\mathrm{pm}$)
along its length and consider three different examples (Fig. \ref{f:sys}a-c).
A uniformly attractive pore, Pore $\alpha$, serves as the control case. Pore $\beta$ is structured
to have an attractive entrance and exit separated by a repulsive core, while Pore $\gamma$ is designed
to have an attractive entrance and a repulsive exit. These apparently minor changes in the pore
patterning turn out to have significant effects on the translocation times.

\begin{figure*}
\includegraphics[width=0.95\linewidth]{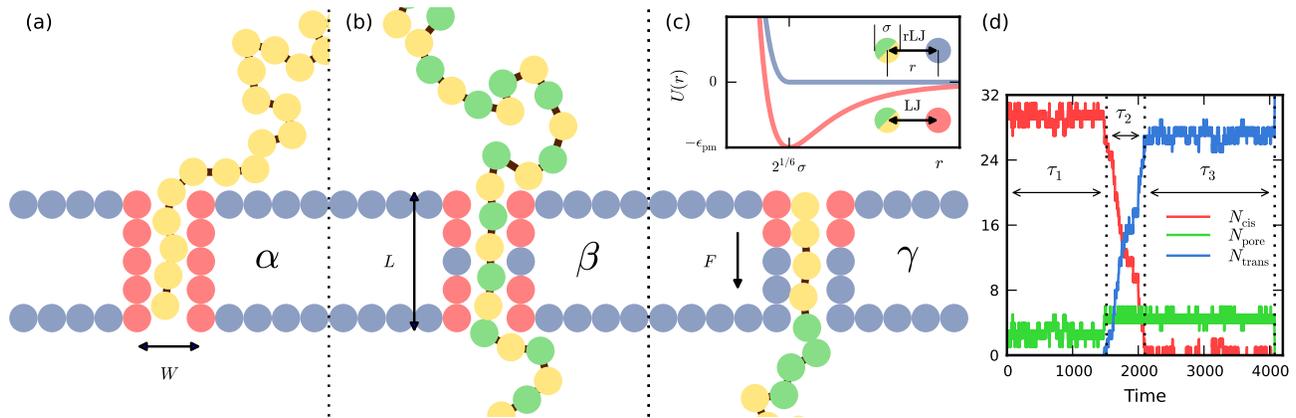}
\caption{\label{f:sys}
{The schematics of the polymer and nanopore models.}
Simulation snapshots showing the translocation of a homopolymer and heteropolymers with
different block lengths across the three patterned pores at various stages of the
translocation process, namely, filling, transfer and escape. (a) A homopolymer (yellow)
translocating from the cis to the trans side of Pore $\alpha$. The interaction of the pore
monomers (red) with the polymer has an attractive well whereas the monomers that make up
the walls of the pore (blue) have an excluded volume interaction with the polymer. The pore
width is fixed and there is a constant force driving the polymer that acts inside the pore.
(b) A heteropolymer (poly(dAdC)$_{16}$) of block length $M=2$ with alternating bases
A (yellow) and C (green) translocating across Pore $\beta$. Pore $\beta$ consists
of two sticky monomers on either end of the pore that are separated by a wall monomer.
The bases A and C have different interactions with the sticky monomers.
(c) A heteropolymer (poly(dA$_4$dC$_4$)$_8$) of block length $M=8$ translocating across
Pore $\gamma$, which has two sticky monomers on the cis side and three repulsive monomers
on the trans side that result in an attractive entrance and a repulsive exit.
(Inset) Shows the pore-polymer potentials. (d) A trace of the monomer count at the trans
end, middle, and cis end of the pore as functions of time for a homopolymer translocating
through Pore $\gamma$, with $\epsilon_{\mathrm{pm}}= 3$ and $F = 0.5$.
$\tau_1$, $\tau_2$ and $\tau_3$ change dramatically when pore patterning is introduced.
}
\end{figure*}

\section{Translocation Time Distributions for Homopolymers}

The translocation time ($\tau$) is divided into
(i) the initial {\em filling} time ($\tau_1$): the time taken by
the first monomer of the polymer to reach the exit without returning to the pore,
(ii) the {\em transfer} time ($\tau_2$): the time taken from the exit of
the first monomer into the trans side to the entry of the last monomer
from the cis side, and (iii) the {\em escape} time ($\tau_3$): the time between entry
of the last monomer in the pore and its escape to the trans side; see Fig. \ref{f:sys}a-c.
These definitions are better characterized by counting the number of monomers of the polymer
on the cis side, $N_\mathrm{cis}$, inside the pore, $N_\mathrm{pore}$, and on the trans
side, $N_\mathrm{trans}$, as functions of time (Fig. \ref{f:sys}d), with
$N = N_\mathrm{cis} + N_\mathrm{pore} + N_\mathrm{trans}$ (see Supplementary Movie 1).

\begin{figure*}[t!]
\includegraphics[width=0.90\linewidth]{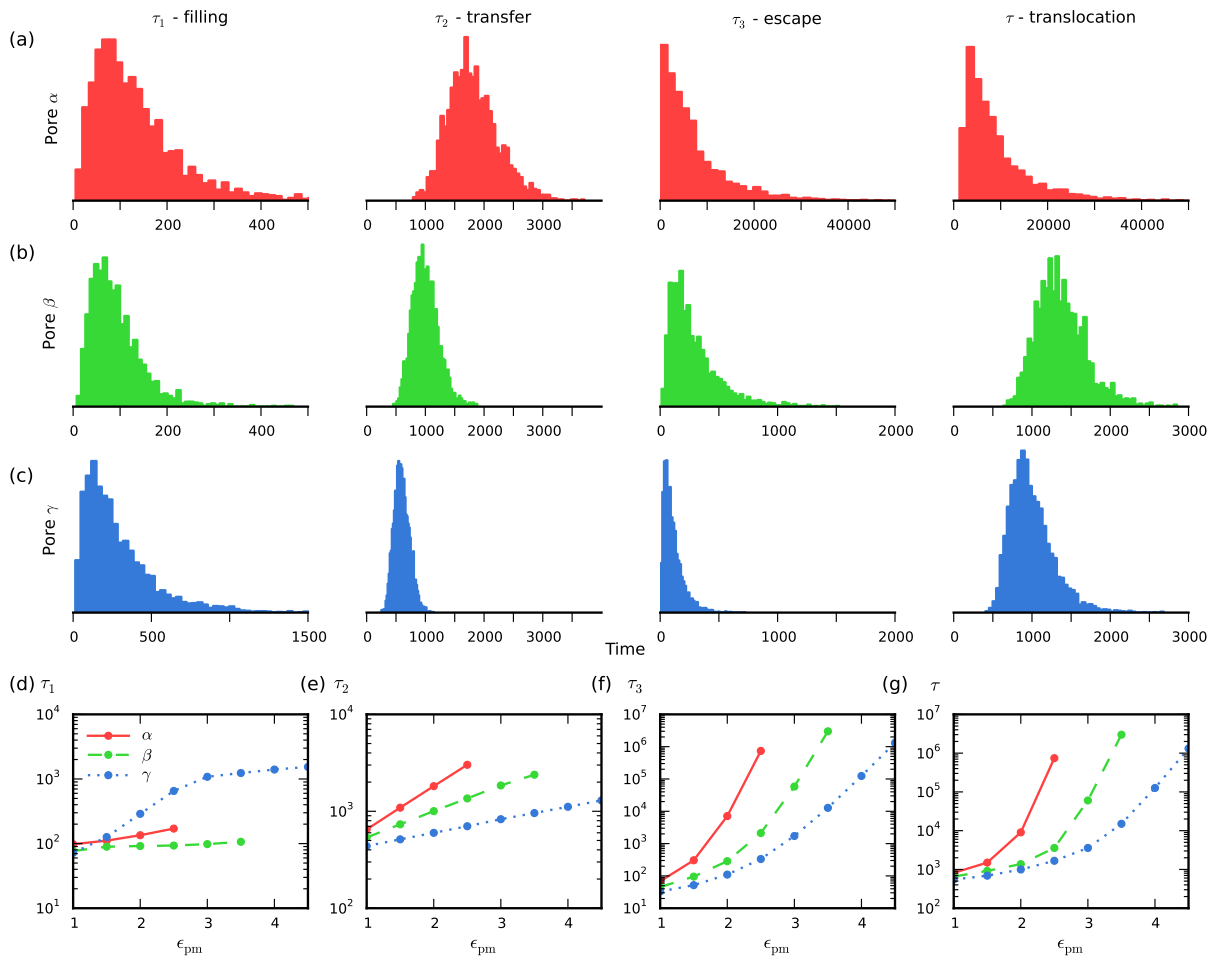}
\caption{{Translocation time statistics for homopolymers.}
(a-c) Comparison of translocation time distributions
for the three patterned pores, for $F=0.5$ and $\epsilon_\mathrm{pm}=2.0$.
The filling, transfer and escape distributions are similar across the three pores,
but have distinctly different scales (e.g. average and variance), such that
the overall translocation time distribution for the three pores are discernably
different. (d-f) Comparison of average filling, transfer and escape times for
the three different pores for $F=0.5$ as a function of $\epsilon_{\mathrm{pm}}$.
While the filling time shows only a moderate dependence on the stickiness, and
the transfer time exhibits an exponential dependence on $\epsilon_{\mathrm{pm}}$,
the dependence of the the escape time is even faster than exponential.
(g) The total translocation time for the three pores as a function of $\epsilon_{\mathrm{pm}}$
indicates that for small forces, the escape time dominates the translocation process.
The orders of magnitude differences in the translocation times between different pores shows the
extraordinary sensitivity of the translocation dynamics on pore patterning.
}
\label{f:homo}
\end{figure*}

\begin{figure*}[t!]
\includegraphics[width=0.95\linewidth]{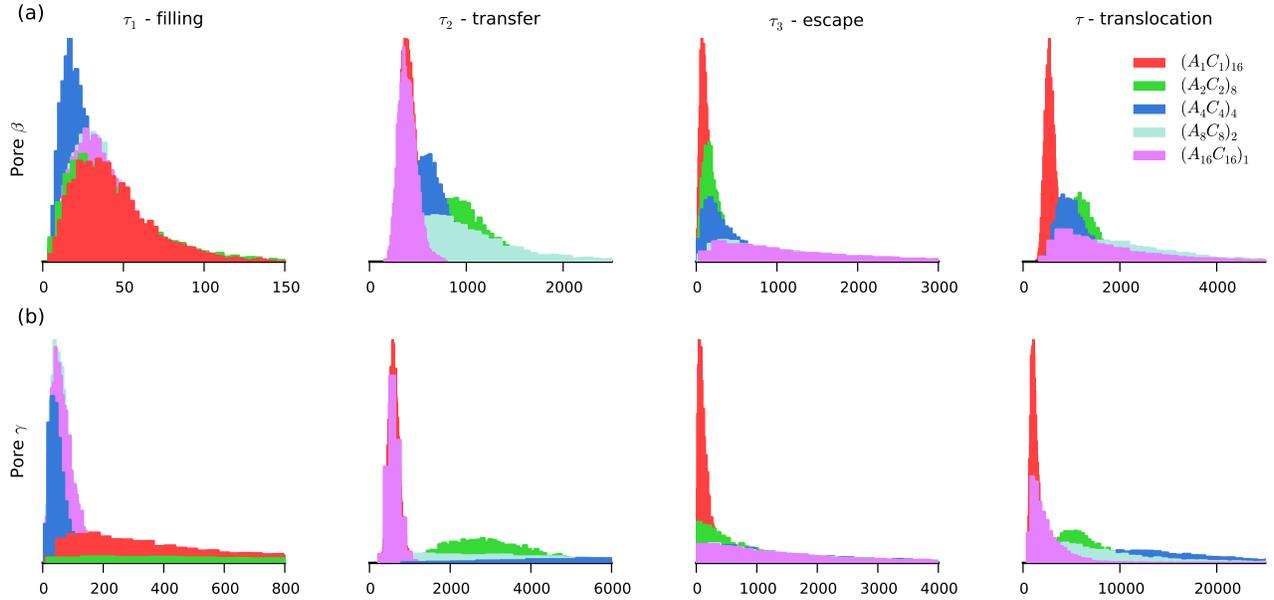}
\caption{{Translocation time statistics for heteropolymers.}
Comparison of filling, transfer, escape and translocation time distributions
for Pores $\beta$ (a) and $\gamma$ (b) and five different sequences of
the heteropolymer. The distributions correspond to $F = 1.0$ for Pore $\beta$,
and $F = 0.5$ for Pore $\gamma$, respectively.
}
\label{f:hetero}
\end{figure*}

For Pores $\alpha$ and $\beta$, the filling time, $\tau_1$, depends weakly on the stickiness
of the pore (Fig. \ref{f:homo}d). In the presence of the weak driving force, Pore $\beta$
has a shallower potential well near the
cis end which reduces trapping time making filling easier (see Appendix B).
The barrier encountered near the core is small enough to be
overcome by the fluctuations of the polymer. With increasing $\epsilon_\mathrm{pm}$,
the effect of trapping becomes more dominant and thus the difference in $\tau_1$ between
pores $\alpha$ and $\beta$ increases. Pore $\gamma$, which has a repulsive exit,
takes a relatively longer time to fill. The large potential barrier beyond the cis side
slows down the polymer increasingly more as the entrance becomes stickier
(with increasing $\epsilon_\mathrm{pm}$). The distribution of filling times
shows a relatively longer exponential tail for Pore $\gamma$ due to this potential barrier.
In sharp contrast to the filling time, the transfer time $\tau_2$ has a much more regular
behavior with increasing stickiness of the pores (Fig. \ref{f:homo}e). The transfer
of the polymer over the length of the pore depends on the potential landscape
inside the pore: Pore $\alpha$, which is attractive throughout, has the longest transfer
time, while Pore $\gamma$, which is the least attractive pore, has the shortest $\tau_2$.
Figure \ref{f:homo}e shows that the transfer time depends exponentially on $\epsilon_\mathrm{pm}$, and
the difference in scales across the three pores is consistent with the number of attractive beads
inside each pore.
The escape time, $\tau_3$, depends strongly on the pore interaction near
the exit, and differs most dramatically across the three pores (Fig. \ref{f:homo}f).
In this time interval, the polymer is already inside the pore and to
escape the pore it needs to overcome the potential barrier near the exit.
The dependence of the escape time on $\epsilon_\mathrm{pm}$ turns out to be faster
than exponential, which suggests that seemingly insignificant changes in the
stickiness pattern and strength of the wall of the pore could modify the
translocation time by several orders of magnitude.

The average total translocation time for the homopolymer across all the
pores is plotted in Fig. \ref{f:homo}g, which shows that for the relatively
weak external force used here the translocation process is controlled by
the escape mode (see Fig. \ref{f:homo}d-g). The overall translocation time distributions
for the three pores are also very different (Fig. \ref{f:homo}a-c), despite
the fact that the general shape of the distributions for each mode of
the translocation process were similar. The extreme sensitivity of
the translocation dynamics of the homopolymer on the pore patterning and stickiness
suggests that it might be possible to engineer pores such that heteropolymers
of any given sequence will have distinct statistical features that could be used
for stochastic sequence detection.

\section{Heteropolymer Sequence Sensing}

To examine the feasibility of this sequencing strategy, we replace the homopolymer
with heteropolymers constructed in accordance with earlier experimental \cite{Meller2000}
and theoretical \cite{luo2008prl} studies of polynucleotide translocation through nanopores;
those containing symmetric purine-pyrimidine blocks of the form $A_nC_n$, with the block
length $M = 2n$ (Fig. \ref{f:sys}; see Appendix A). We assign different values to the
attractive interactions of the sticky beads in the pore with base A ($\epsilon_\mathrm{pA}$)
and base C ($\epsilon_\mathrm{pC}$), with $\epsilon_\mathrm{pA} > \epsilon_\mathrm{pC}$.

\begin{figure*}[t!]
\includegraphics[width=0.95\linewidth]{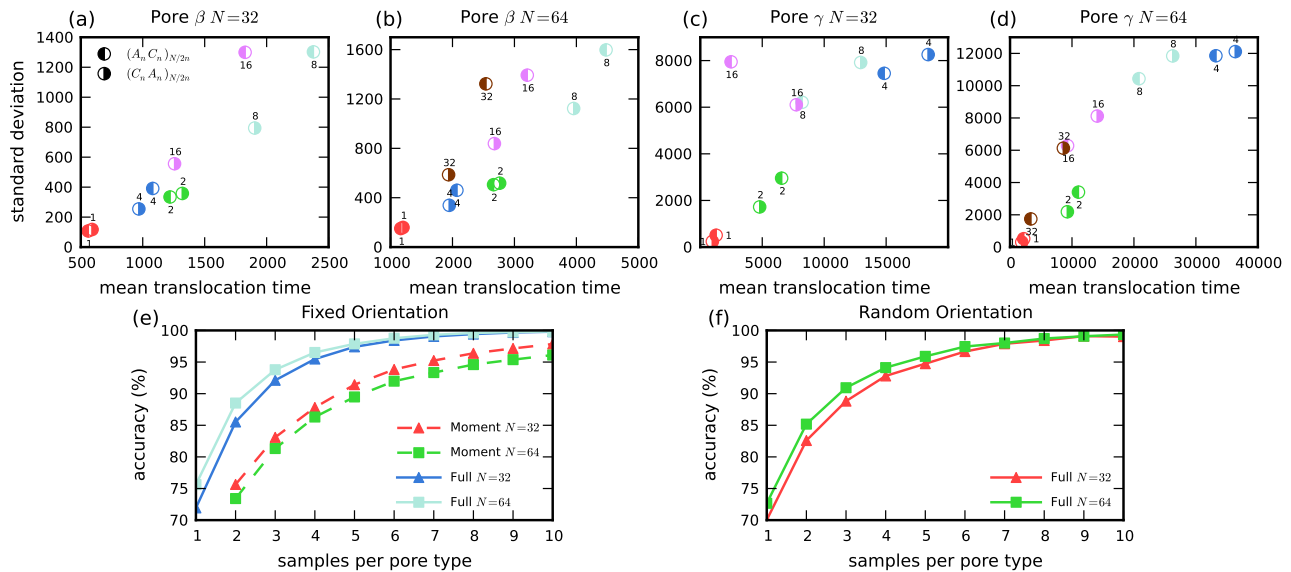}
\caption{{Using translocation time statistics to detect polynucleotide sequences.}
(a-d) Scatter plots showing the distinctive mean and standard
deviations of the different sequences and for different ends of the
polynucleotide entering the pore. The mean translocation time,
$\langle \tau \rangle$, and its standard deviation,
$\sqrt{\langle \tau^2 \rangle - \langle \tau \rangle^2}$, for the different
sequences are calculated from the distributions. The scatter plots reveal the distinctive
characteristics of the translocation events for the different sequences through each pore.
The plots correspond to
(a) Pore $\beta$ with $F = 1.0$ and $N = 32$
(b) Pore $\beta$ with $F = 1.0$ and $N = 64$
(c) Pore $\gamma$ with $F = 0.5$ and $N = 32$ and
(d) Pore $\gamma$ with $F = 0.5$ and $N = 64$.
(e-f) Accuracy of sequence detection using multiple joint translocation events through
Pores $\beta$ and $\gamma$. The plots are constructed by recording a given number of translocation
times through Pores $\beta$ and $\gamma$, and using a comparison with either the full distribution
or the first two moments of the distribution shown in the scatter plots (a-d). For the method
that uses the moments, the sample average and standard deviation are calculated and used to find
the relative error of the sample average and standard deviation compared to the known
values and compound them into an error metric for each pore and each sequence. The error metric
is subsequently used for each pore to define a closeness metric, which will be minimized
to predict the sequence. The accuracy is the ratio between the number of successful
predictions and the total number of attempts.
In (e) the orientation of the polymer is known and preserved when it passes through the different pores.
In (f) the orientation of the polymer is not known and randomly changes when it passes through the different pores.
}
\label{f:orient}
\end{figure*}

The translocation time distributions for five different sequences are shown in
Fig. \ref{f:hetero}a-b for Pores $\beta$ and $\gamma$. We find that the different modes
of translocation across the two pores respond differently to variations in the block length,
such that the outcome for the total translocation time exhibits distinct features (see
Supplementary Movies 2 and 3 and Appendix C). To simplify the picture, we summarize the distributions
for each pore in a scatter plot by using only the two basic characteristics of mean and
standard deviation (Fig. \ref{f:orient}a-d). We observe a number of interesting features.
For example, both mean time and standard deviation seem to roughly increase with block
length for Pore $\beta$, whereas for Pore $\gamma$ mean time initially increases
with block length, peaks at $n=4$ and goes back to smaller values for longer blocks.
While Pore $\beta$ cannot easily distinguish between $(A_4C_4)_4$ and $(A_2C_2)_8$,
Pore $\gamma$ can, and the reverse is true for $(A_4C_4)_4$ and $(A_8C_8)_2$. We have also
examined the effect of the orientation of the heteropolymer when it enters the pore, and
considered polymers of total length $N=32$ and $N=64$ (Fig. \ref{f:orient}a-d). The
differences in the scatter immediately suggests that a combined translocation time
measurement across the two pores and comparison with the statistics of the known
sequences could help identify an unknown sequence to a high accuracy.

To demonstrate this idea and probe its statistical robustness as a sequencing
strategy, we run a test on a model sequencing device that would be made up of
multiple copies of Pores $\beta$ and $\gamma$ that are arranged in-series, such that
the readout from translocation of a given polynucleotide with an unknown sequence
through all of them can be independently recorded. We calculate the average and
standard deviation of the translocation times through the Pores $\beta$ and $\gamma$,
separately, using their corresponding multiple readouts. Using the difference between
the measured means and standard deviations and the tabulated values for known sequences
through each pore, we calculate the relative error for each sequence and minimize it
for all sequences across both pores to find the closest match, which will be returned
as the predicted sequence. The ratio of the number of successful sequence detection events
and the total number of attempts, which is defined as the accuracy of the statistical
sequence detection algorithm, turns out to be remarkably high (Fig. \ref{f:orient}e).
For $N=32$ and fixed orientation of the polynucleotide for all pores, the accuracy
starts off at $75\%$ with just the minimum two copies of each pore
and rises quickly to above $95\%$ when there are ten copies of each pore.

Instead of just using the first two moments, we can choose to use the full
translocation time distributions for the sequence detection, using
the following method. If we make a measurement of the
translocation time ($\tau$) of a polymer with an unknown sequence through a
given pore (Pore $\beta$, say), then the probability of the time being
part of a distribution of a known sequence (say $n$) is ${\cal{P}}_n^\beta(\tau)$,
where ${\cal{P}}_n$ is the known probability distribution.
After $m$ measurements, the likelihood of the translocation times being
part of a given distribution can be defined as
${\cal{L}}_n = \Pi_{i=1}^m {\cal{P}}_n^\beta(\tau_i)$. The structure of the unknown
heteropolymer is determined by finding the $n$ with the maximum likelihood,
$\mathrm{seq} \equiv \mathrm{seq}[\mathrm{max}\{{\cal{L}}_n\}]$. For multiple pores
and fixed orientation of the polymer through all of them, the likelihood can be generalized to
${\cal{L}}^{\rm fo}_n =\Pi_{i=1}^m {\cal{P}}_n^\beta(\tau_i) \Pi_{i=1}^m {\cal{P}}_n^\gamma(\tau_i) ...$.
Figure \ref{f:orient}e shows the resulting accuracy plots as obtained using the full
translocation time distributions, which exhibit a considerably faster convergence
in the algorithm. The corresponding results are very similar for $N=64$
(Fig. \ref{f:orient}e). This shows that an inherently statistical DNA sequencing
strategy could be designed to have an arbitrary accuracy.

During the sequence detection process the heteropolymer could enter the pore with
either base A or base C entering first. Therefore it is imperative to consider
orientation effects on the translocation time distributions, as seen in
Fig. \ref{f:orient}a-d, and hence on our sequencing strategy.
Due to the possibility of orientation flips during multiple readouts of the
unknown sequence, we need to consider all permutations of the two
orientations in a given set of readouts. When we incorporate the possibility
of different orientation in the translocation time measurements, then
we would need to consider the sum of all possible permutations of orientations
in determining the likelihood of the translocation times being
part of a given distribution. This leads to
\begin{math}\displaystyle
{\cal{L}}^{\rm ro}_n = \left[{\sum_{\mathrm{perm}}}\Pi_{i=1}^m {\cal{P}}_n^\beta(\tau_i) \right]
\left[{\sum_{\mathrm{perm}}} \Pi_{i=1}^m {\cal{P}}_n^\gamma(\tau_i) \right] \cdots
\end{math}
for the random orientation case.
Again, we count the number of successful sequence detection events and
plot the accuracy of this detection algorithm, as shown in Fig \ref{f:orient}f.
The result of this calculation shows that using the full distribution is surprisingly
robust with respect to the randomization of the orientation, which is of paramount importance in practice.
For the set of measurements which do not involve orientation effects we observe a distinctly faster
detection of a sequence ($95\%$ when there are $4$ copies of each pore)
as compared to the detection using the scatter plots (Fig. \ref{f:orient}e).
With the orientations of the polymer as it enters the pore taken into account, the accuracy
of detection rises to $~ 95\%$ with just $5$ copies of each pore (Fig. \ref{f:orient}f).

\section{Conclusion}

In contrast to the generally accepted notion of suppressing the stochastic element of
polynucleotide motion through nanopores to achieve efficient DNA sequencing,
we propose to extract information from the statistical fluctuations towards
sequence detection. Our strategy is based on designing distinguishable
translocation time statistics for any given sequence by engineering
the polymer-pore interactions and combining readouts from multiple pores
for rapid convergence. The desired patterns in surface interaction could be
achieved by using biological nanopores with appropriate modification \cite{Li2011,Hammerstein2011}
or those with known hydrophobic-hydrophilic pattern structure \cite{Mahfoud2006},
as well as solid-state nanopores with tailor-made surface interactions \cite{Storm2003,Kim2006,Ohshiro2006,Iqbal2007,Wanunu2007}. The proposed approach could potentially improve the overall speed of sequence detection by orders of
magnitude, and could be integrated in high throughput microfluidic devices.

\acknowledgments
We would like to thank G. Battaglia for fruitful discussions.
This work was supported by grant EP/G062137/1 from the EPSRC.

\appendix
\section{Methods}

\begin{figure*}[t!]
\includegraphics[width=0.9\linewidth]{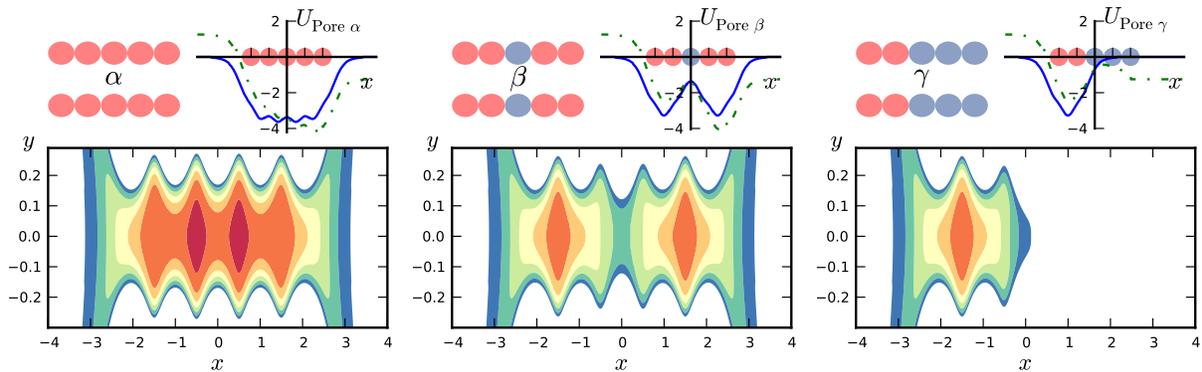}
\caption{\label{f:pc}
Comparison of the polymer-pore interaction potentials.
(Top left) A schematic of the pore. Pore monomers could either have an
attractive (LJ) interaction (red) or a short range repulsive (rLJ) interaction
(blue) with the polymer inside the pore.
(Top right) The potential energy landscape
in the center ($y = 0$) along the length of the channel (blue) is modified
(green) in the presence of an external driving force $F = 0.5$.
(Bottom) The complete potential energy landscape experienced by the polymer
inside the pore. Blue to red represents increasing potential depth.}
\end{figure*}

{\em Homopolymer model.}
We model the polymer as a self avoiding chain by using beads
and springs (Fig. \ref{f:sys}). The beads represent monomer groups of the polymer
and we model the excluded volume interaction between a pair of monomers by
a truncated repulsive Lennard-Jones (rLJ) potential of the form
\begin{eqnarray}
\label{e:lj}
    U^\mathrm{LJ}_\mathrm{mm}(r) &=& \left\{
     \begin{array}{lr}
        4\epsilon \left[ \left(\frac{\sigma}{r} \right)^{12} -  \left(\frac{\sigma}{r}\right)^{6} \right]  + \epsilon &: r \le r _\mathrm{min} \\
       0&: r > r_\mathrm{min}
     \end{array}
   \right.
   \nonumber
\end{eqnarray}
where $\epsilon$ is the potential depth and $\sigma$ is the monomer diameter.
The cut-off distance, $r_\mathrm{min} = 2^{1/6}\sigma$, is set at the potential
minimum. The bonding springs between monomer groups are modelled by a finite
extension non-linear elastic (FENE) potential of the form
\begin{equation}
    U^\mathrm{FENE}_\mathrm{ch}(r) = -\frac{1}{2} k R^2 \mathrm{ln} \left(1-\frac{r^2}{R^2} \right)
\nonumber
\end{equation}
where $k = 7\epsilon/\sigma^2$ and $R = 2\sigma$ are the spring constant and
bond length respectively. FENE potentials are convenient as the bond length
effectively sets the maximum allowed separation between monomer groups.
We use polymers of length $N=32$ and $N=64$ in our simulations.

{\em Heteropolymer model.}
We model the heteropolymers similarly using beads and springs (Fig. \ref{f:sys}) with
the polymer beads representing the bases A and C arranged in symmetric blocks $A_nC_n$.
With a DNA of length $N = 32$, the minimum value of $n = 1$ is for poly(dAdC)$_{16}$
and the maximum value of $n = N/2$ for poly(dA$_{16}$dC$_{16}$). The bases A and C
are only distinguished by their relative interactions with the pore.

{\em Pore model.}
The pore and wall are constructed from stationary monomers separated by a
distance of $\sigma$ from each other. The pore is made up of two rows of
monomers symmetric about the coordinate system with a length $L=5\sigma$
and separated by a width of $W=2.25\sigma$. The pore width is chosen to allow
only single file translocation of the polymer and avoid hair-pin
configurations. The polymer translocates from the cis (entrance) end to
the trans (exit) end of the pore (Fig. \ref{f:sys}). The walls of the pore
extend in the y direction.

{\em Polymer-pore interaction.}
The interaction of the pore with the polymer is tuned such that the interaction
varies along the length of the pore. This interaction
could either be the short-range repulsive form described above
or the standard LJ form:
\begin{eqnarray}
\label{e:lj}
    U^\mathrm{LJ}(r) &=& \left\{
     \begin{array}{lr}
        4\epsilon_\mathrm{pm} \left[ \left(\frac{\sigma}{r} \right)^{12} -  \left(\frac{\sigma}{r}\right)^{6} \right] &: r \le r _c \\
       0&: r > r_c
     \end{array}
   \right.
  \nonumber
\end{eqnarray}
with $\epsilon_\mathrm{pm}$ denoting the potential depth and $r_c = 2.5\sigma$
denoting the cut-off distance. We choose three different pore patterns with
the patterning symmetric about the x-axis:
(1) Pore $\alpha$ is an attractive pore with all the monomers of the pore interacting
with the polymer by the LJ potential. (2) Pore $\beta$ has an attractive entrance
and exit with the first two monomers and the last two monomers of the pore
interacting with the polymer by the LJ potential and the middle monomer being
repulsive. (3) Pore $\gamma$ has an attractive entrance (first two monomers attractive)
and a repulsive exit (last three monomers repulsive). Note that in all the three cases
the pore entrance is chosen to be attractive to successfully initiate translocation.
The stickiness of the pore ($\epsilon_\mathrm{pm}$) is varied during homopolymer
translocation. During the translocation of the heteropolymer the stickiness differs
for base A ($\epsilon_\mathrm{pA}$) and base C ($\epsilon_\mathrm{pC}$). We fix these
values to $\epsilon_\mathrm{pA} = 3.0$ and $\epsilon_\mathrm{pC} = 1.0$ respectively.
The polymer interacts with the wall
($U_\mathrm{mw}^\mathrm{LJ}$) with the same rLJ potential as used for the
intra-monomer excluded volume interaction. In addition the polymer experiences
a driving force, $\mathbf{F}_\mathrm{e} = F\hat{\mathbf{x}}$
directed along the pore axis with magnitude $F$, which mimics the
electrophoretic driving of biopolymers through nanopores.

{\em Polymer injection.}
In our simulation we are not concerned with injection of the polymer into the
pore, but only with the dynamics of the polymer during translocation. We initially
place the first bead of the polymer chain at the entrance of the pore
and allow the remaining beads to fluctuate. Once the polymer relaxes
to its equilibrium configuration, the bead is released and the translocation
of the polymer across the pore is monitored. The translocation time is defined
as the time that elapses between the entrance of the first bead of the polymer
and the exit of the last bead. All failed translocation events are discarded.

{\em Integration algorithm.}
The equations of motion of the monomers of the polymer were integrated
using a Langevin dynamics (LD) algorithm that includes a velocity Verlet
update \cite{allen}.
Within the LD formalism, the interaction of the monomers with a solvent is
simulated by a viscous drag term proportional to the monomer velocity and a
random force term modeled by Gaussian white noise with an auto-correlation
function that satisfies the fluctuation-dissipation theorem. The equation of
motion for a monomer therefore takes the form:
\begin{equation}
m{\bf \ddot{r}}_i=- {\mbox{\boldmath$\nabla$}} U_i + \mathbf{F}_\mathrm{ext} - \zeta {\bf v}_i + {\mbox{\boldmath$\eta$}}_{i},
\nonumber
\end{equation}
where $m$ is the monomer mass, $U_i = U_\mathrm{mm}^\mathrm{LJ} + U_\mathrm{ch}^\mathrm{FENE} + U_\mathrm{wm}^\mathrm{LJ} + U_\mathrm{pm}^\mathrm{LJ}$ is the
total potential experienced by a monomer, $\zeta$ is the friction coefficient,
${\bf v}_i$ is the monomer velocity, and ${\mbox{\boldmath$\eta$}}_i$ is the
random force with
$\langle {\mbox{\boldmath$\eta$}}_{i}(t) \cdot {\mbox{\boldmath$\eta$}}_{j}(t_0) \rangle
=4 k_{\rm B} T \zeta \delta_{ij}\delta (t-t_0)$, $T$ being the temperature.
A time step of $\Delta t = 0.01$ is used in all simulation runs.

\begin{figure}
	\includegraphics[width=0.95\linewidth]{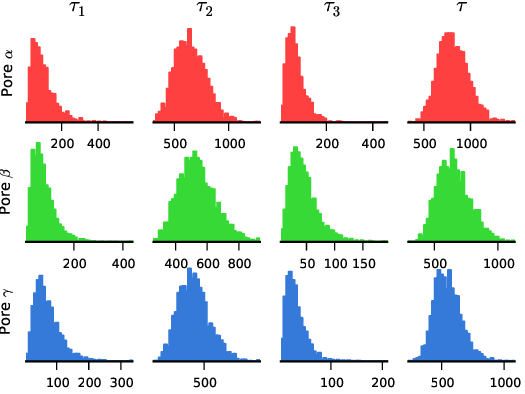}
	\caption{\label{f:homo1}
Comparison of the filling, transfer, escape and translocation time distributions
for the three pores at $F = 0.5$ and $\epsilon = 1.0$
}
\end{figure}

\begin{figure}
	\includegraphics[width=0.95\linewidth]{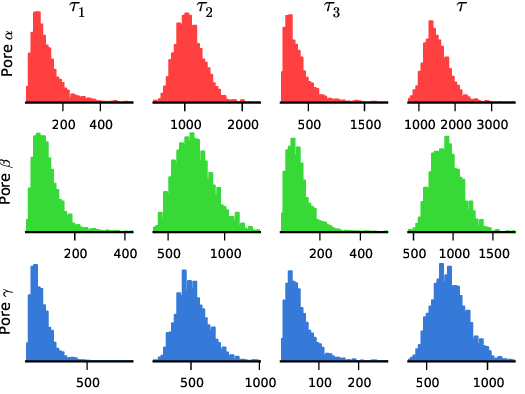}
	\caption{\label{f:homo2}
Comparison of the filling, transfer, escape and translocation time distributions
for the three pores at $F = 0.5$ and $\epsilon = 1.5$
}
\end{figure}

\begin{figure}[]
	\includegraphics[width=0.95\linewidth]{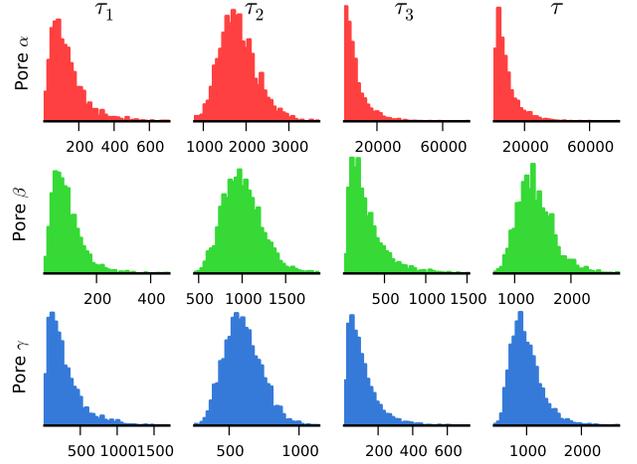}
	\caption{\label{f:homo3}
Comparison of the filling, transfer, escape and translocation time distributions
for the three pores at $F = 0.5$ and $\epsilon = 2.0$
}
\end{figure}

{\em Reduced units.}
The units of energy, length, and mass are set by $\epsilon$, $\sigma$, and $m$,
respectively. These set the scale for the time as $(m\sigma^2/\epsilon)^{1/2}$.
Following Luo et al. \cite{luo2008prl}, we assume that the size of each bead
in our coarse-grained polymer model corresponds to the Kuhn length of
a single stranded DNA, which is approximately three nucleotide bases. This sets the
bead size, $\sigma \approx 1.5$\,nm, the mass of the bead, $m \approx 936$\,amu
(given that the mass of a base in DNA is $\approx 312$\,amu) and the charge of a bead,
$q \approx 0.3$\,e (each base having a charge of $0.1$\,e effectively \cite{Sauer2003}).
We set $\zeta = 0.7$ and $k_BT = 1.2$ to allow comparison with known results.
Therefore, the interaction strength at $T = 295$\,K is given by $\epsilon = k_BT/1.2
\approx 3.4 \times 10^{-21}$\,J. This gives the time scale of
$(m\sigma^2/\epsilon)^{1/2} \approx 30$\,ps and a force scale of
$\epsilon/\sigma \approx 2.3$\,pN. Therefore an external driving force in the range
$0.5 - 1.0$ corresponds to a voltage range $V = FL/q \approx 190-380$\,mV across the
pores. Note, however, that higher values of up to $0.5$\,e for the effective base charge
have also been reported in the literature \cite{keyser2006}, which suggest that
the appropriate voltage range could be lower than the above-mentioned values.
As a rough indication of how much the patterning could affect translocation speed,
we note the example of a homopolymer with $\epsilon_\mathrm{pm} = 2.5$
(see Fig. \ref{f:homo}g), which yields the translocation time of $100$ $\mu$s for
Pore $\alpha$ and $0.1$ $\mu$s for Pore $\gamma$. These figures are
consistent with the typical \cite{Branton2008} observed translocation rates
of a single nucleotide$/\mu$s.

\begin{figure}[]
	\includegraphics[width=0.95\linewidth]{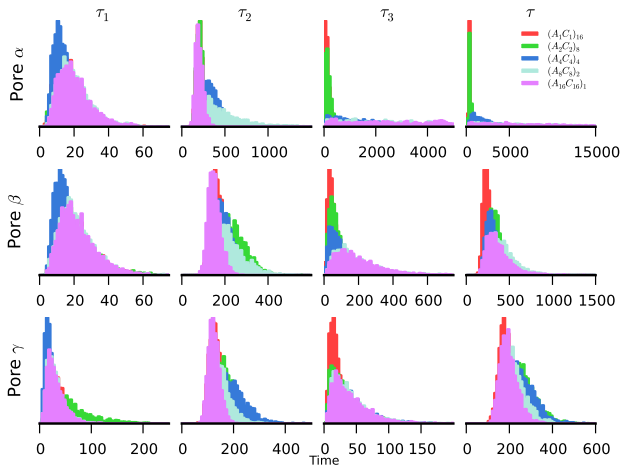}
	\caption{\label{f:hetero1}
Comparison of the filling, transfer, escape and translocation time distributions
for the three pores at $F = 1.0$ and $W = 2.5$ using heteropolymers with different sequences.
}
\end{figure}

\begin{figure}[]
	\includegraphics[width=0.95\linewidth]{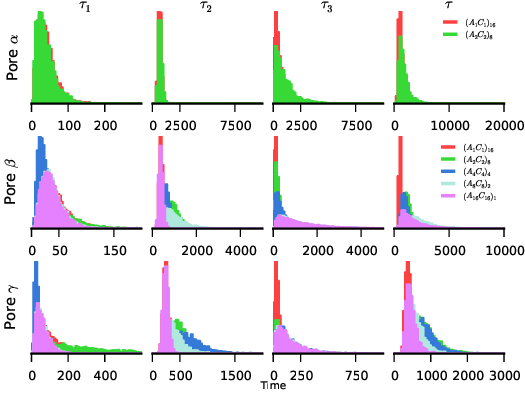}
	\caption{\label{f:hetero2}
Comparison of the filling, transfer, escape and translocation time distributions
for the three pores at $F = 1.0$ and $W = 2.25$ using heteropolymers with different sequences.
}
\end{figure}

\begin{figure}[]
	\includegraphics[width=0.95\linewidth]{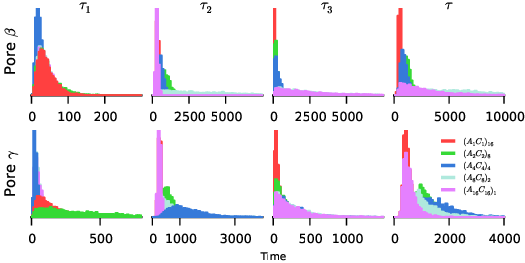}
	\caption{\label{f:hetero3}
Comparison of the filling, transfer, escape and translocation time distributions
for the Pores $\beta$ and $\gamma$ at $F = 0.5$ and $W = 2.5$ using heteropolymers with different sequences. Translocation through Pore $\alpha$ is extremely slow for these range of
values and are not considered for this analysis.
}
\end{figure}
\begin{figure}[]
	\includegraphics[width=0.95\linewidth]{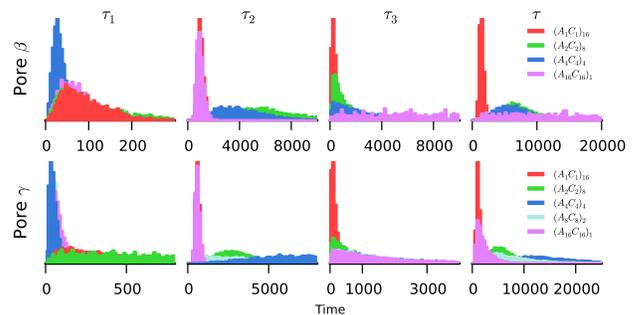}
	\caption{\label{f:hetero4}
Comparison of the filling, transfer, escape and translocation time distributions
for the Pores $\beta$ and $\gamma$ at $F = 0.5$ and $W = 2.25$ using heteropolymers with different sequences. Translocation through Pore $\alpha$ is extremely slow for these range of
values and are not considered for this analysis.
}
\end{figure}

\section{Sensitivity of translocation time distributions on pore-polymer interaction}

The translocation of a homopolymer through a narrow pore has been theoretically studied extensively in the past \cite{lubensky,luo2006jcp,muthukumar,matysiak,luo2007prl,luo2008pre,luo2008prl,gauthier,luan,nikoubashman,sung,muthukumar1999,muthukumar2001,kantor2001,muthukumar2003,metzler,slonkina,kantor2004,milchev2004,gerland,gopinathan,wong,milchev2011,abdolvahab2011}.
Luo et al. \cite{luo2006jcp,luo2007prl,luo2008pre} investigated the pore-polymer interaction in
a uniformly attractive pore, which we chose as the control pore, Pore $\alpha$, in our simulations.
In Fig. \ref{f:pc}, we compare the potential landscapes for Pores $\alpha$, $\beta$, and $\gamma$,
respectively. The potential landscapes seen by the polymer at the center of the pore along
its length reveal the positions of the potential barriers for the three pores. Pore $\alpha$ has a strong
barrier at the exit due to the stickiness of the pore. Pore $\beta$---that has a repulsive part in the
middle---has a far lower barrier at the exit. It does, however, experience a small barrier just
after the entrance, which it overcomes easily. This explains why the filling time for Pore $\beta$
is lower than that of Pore $\alpha$. Pore $\gamma$, which has an attractive entrance but a repulsive exit,
has a large barrier at the entrance, which makes the filling time relatively longer as compared the other pores.
However, the escape time for the polymer in Pore $\gamma$ is vastly reduced due to the repulsive exit.

In Figs. \ref{f:homo1}, \ref{f:homo2}, and \ref{f:homo3}, we observe the change of the distribution for the
translocation times with increasing attractive strengths, $\epsilon_\mathrm{pm}$ for the the three
different pores. For Pore $\alpha$, we note the transition from a Gaussian form to a long-tailed
distribution with increasing attraction, which was observed by Luo et al. This transition is observed
in all the three pores, although for Pores $\beta$ and $\gamma$ they happen at higher attractive
strengths.

\section{Optimizing the pore for sequencing: the effect of driving force and pore width}

To understand the effects of patterning the pore on sequencing,
we considered the translocation of heteropolymers through the pores. Following
Luo et al. \cite{luo2008prl}, the polymers were represented as consisting of symmetric blocks
$A_nC_n$ of A and C bases, which interact differently with the pore. The time distributions
for the three pores show a varying degree of sensitivity on the specific sequence of the polymer
(Figs. \ref{f:hetero1}, \ref{f:hetero2}, \ref{f:hetero3}, and \ref{f:hetero4}), depending on
the strength of the external force and the pore width. In Fig. \ref{f:hetero1},
we show the dependence of the distributions for $F = 1.0$ and $W = 2.5$. For Pore $\alpha$,
the difference in distributions for short block lengths is relatively small. As the block lengths
are increased, the distribution changes sharply. However, for larger block lengths it becomes
difficult again to distinguish them. For Pores $\beta$ and $\gamma$, the distributions
have a high degree of overlap and are not suitable for sequencing.

As the width is decreased (Fig. \ref{f:hetero2}, $W = 2.25$), Pore $\alpha$ takes extremely long to
translocate for larger block lengths. The potential barrier proves difficult to surmount and
the polymer is stuck for long periods inside the pore. However, lowering the width has a positive
impact on Pore $\beta$ which leads to translocation time distributions that can distinguished from
one another. Although the translocation times are much longer, the distributions are well separated
by their means and standard deviations. This impact is much less for Pore $\gamma$.

On the other hand, we could keep the pore width fixed ($W = 2.5$) and lower the strength of
the external driving force (Fig. \ref{f:hetero3}). The effect on Pore $\alpha$ is drastic
as the polymers fail to cross the potential barrier. Pore $\beta$ and $\gamma$, on the other hand,
still translocate polymers although their distributions for the different sequences are far
from distinguishable.

Finally, we keep the width at $W = 2.25$ and lower the force to $F=0.5$ (Fig. \ref{f:hetero4}).
This completely takes out Pore $\alpha$ from consideration as the translocation times become
prohibitively long. The translocation time scales are now much longer for Pores $\beta$ and
$\gamma$ as well. However, the mean and standard deviations for Pore $\gamma$ are again well
separated making it easier to distinguish between the distributions, and hence make it a suitable
candidate for sequencing. In our simulations, we use Pore $\beta$ at $F = 1.0, W = 2.25$ and
Pore $\gamma$ at $F = 0.5, W = 2.25$ as the two most suitable pores for our sequencer.

\end{document}